\documentclass[%
 reprint,
 amsmath,amssymb,
 aps,
]{revtex4-2}

\usepackage[usenames]{color}
\usepackage{colortbl}
\usepackage{graphicx}
\usepackage{dcolumn}
\usepackage{bm}

\begin{document}

\preprint{APS/123-QED}

\title{Cascade Brillouin scattering on short-lived phonons
for flat-top frequency comb generation}

\author{Egor R. Verevkin}
\affiliation{Moscow Center for Advanced Studies, 20 Kulakova, Moscow, Russia}
\author{Ilya V. Doronin}
\affiliation{Dukhov Research Institute of Automatics (VNIIA), 22 Sushchevskaya, Moscow, Russia}
\affiliation{Moscow Center for Advanced Studies, 20 Kulakova, Moscow, Russia}
\affiliation{Institute for Theoretical and Applied Electromagnetics, 125412, 13 Izhorskaya, Moscow, Russia}
\author{Alexander A. Zyablovsky}
\email{zyablovskiy@mail.ru}
\affiliation{Dukhov Research Institute of Automatics (VNIIA), 22 Sushchevskaya, Moscow, Russia}
\affiliation{Moscow Center for Advanced Studies, 20 Kulakova, Moscow, Russia}
\affiliation{Institute for Theoretical and Applied Electromagnetics, 125412, 13 Izhorskaya, Moscow, Russia}
\author{Evgeny S. Andrianov}
\affiliation{Dukhov Research Institute of Automatics (VNIIA), 22 Sushchevskaya, Moscow, Russia}
\affiliation{Moscow Center for Advanced Studies, 20 Kulakova, Moscow, Russia}
\affiliation{Institute for Theoretical and Applied Electromagnetics, 125412, 13 Izhorskaya, Moscow, Russia}
\date{\today}




\begin{abstract}
We consider Brillouin scattering on short-lived phonon modes confined within a subsection of optical ring resonator. Such that a) the relative Brillouin shift between propagating and scattered waves is smaller than the relative width of phonon modes, b) one phonon mode provides large integrated overlap with many pairs of photon modes. In this case one phonon mode facilitates scattering between many pairs of optical modes. We show that in this limit two phonon modes are sufficient for cascade Brillouin scattering (one forward propagating wave and one counter propagating wave), and that the cascade behavior is qualitatively different from the cascade in conventional Brillouin systems with distinct phonon modes for each optical mode pair. In particular, our results show that there is a pump threshold above which many optical modes become excited simultaneously, as opposed to a cascade gradually building up. The resulting cascade scattering can be exploited for flat-top frequency comb generation.
\end{abstract}

\maketitle
\section*{Introduction}
Brillouin scattering is light scattering on acoustic waves, i.e., on phonons \cite{wolff2021brillouin}, caused by electrostriction effect \cite{newnham1997electrostriction}. This can take the form of optomechanical instability - when light in medium creates phonon wave via electrostriction \cite{beugnot2012electrostriction} that consequently scatter light, creating new output spectral peaks. This phenomenon has been used to generate coherent light in so-called Brillouin laser \cite{otterstrom2018silicon}. Brillouin laser consists of a cavity, typically a waveguide or a ring resonator, pumped via a waveguide with external light. The optomechanical instability results in generation of light waves with lower frequency (Stokes component) than the pump light \cite{wolff2021brillouin}. The frequency shift is determined by phonon properties.

Coupling waveguide to the resonator allows this generated light to be harnessed and used.  The very same waveguide that facilitates pumping usually doubles as the waveguide for output field \cite{otterstrom2018silicon}.  Notably, the light wave generated in such a manner can give rise to another phonon wave and scatter on it. Thus, a cascade of Brillouin scattering can occur in one device \cite{behunin2018fundamental, jin2023intrinsic}. The model currently used to describe cascaded Brillouin scattering considers a series of optical modes and a series of phonon modes \cite{behunin2018fundamental}. In this model each phonon mode is responsible for Brillouin scattering from a single light mode propagating in one direction to a single light mode propagating in the opposite direction.

One of the promising applications of cascaded Brillouin scattering is frequency comb generation. Frequency comb is a signal with the spectrum comprised of regularly spaced lines \cite{picque2019frequency}. Cascaded Brillouin scattering generates spectral lines with higher wavelength than pump wavelength \cite{behunin2018fundamental}. A downside of cascaded Brillouin scattering for comb generation is that spectral lines are not perfectly equidistant due to photons losing a fraction of their current energy upon scattering, rather than losing a constant amount of energy \cite{behunin2018fundamental}. The amplitudes of Brillouin lines are also irregular. Irregularity in comb structure limits its usefulness for frequency measurements since the protocols require equidistant lines \cite{udem2002optical}, however, other potential applications of frequency combs, such as in optical communication \cite{choudhary2017chip, pelusi2022brillouin} or optical computation \cite{okawachi2023chip, sun2023applications}, can make use of it, since the spectrum, although not evenly spaced and uniform, has predictable shape. Currently, cascaded Brillouin scattering is successfully used in experiments for frequency comb generation \cite{redding2022high,maksymov2022acoustic}, especially in tandem with other nonlinear effects in the medium \cite{huang2019temporal, bunel2025brillouin, pang2023frequency, zhang2024broadband}.

Among methods for generating frequency combs are Kerr nonlinearity and electro-optical modulation. These methods are capable of providing broadband frequency combs  \cite{zhang2024advances, herr2012universal, lugiato2018lugiato, chembo2013spatiotemporal,levy2010cmos} with significantly more lines than Brillouin scattering that are evenly spaced. However, the amplitudes remain non-uniform \cite{he2016dynamics, campbell2024frequency, wang2023flat} and require complex setups to achieve uniformity of 0.7-1 dB, including additional filtering, stabilization~\cite{zeng2026integrated}, or second and third order dispersion engineering~\cite{li2025ultra}. Uniform frequency combs are a crucial component for analogue-to-digital conversion \cite{fang2025320, zazzi2020fundamental}, and as such, search for new tools and methods for generating such combs continues

In this paper, we suggest an approach to generate a uniform frequency comb via Brillouin scattering. We propose a system where many optical modes are pumped via only two phonon modes, unlike previously reported systems where Brillouin scattering occurs via cascade of phonon modes. The two-phonon-mode excitation we suggest is possible for systems with a large ratio $\Omega_j : \Gamma_j$, reaching up to 100:1 in some systems \cite{lu2016stimulated, gundavarapu2019sub, otterstrom2018silicon}. Consequently, the phonon linewidth is sufficiently broad to cover the frequencies of multiple high-order Stokes shifts. In addition to a broad linewidth, our approach requires that the wavevector-matching condition is satisfied approximately for multiple photon mode pairs. This can be achieved by limiting the region of interaction of optical and phonon modes in the resonator. For such a system, we predict that there is a collective pump threshold where all higher-order optical modes ($n > 2$) are amplified simultaneously. The simultaneous appearance of many orders of optical modes enables generation of frequency combs with uniform integrated power per line in the co-propagating comb (flatness below 0.002 dB), which are of interest to signal processing applications, specifically, for analogue-to-digital conversion.

\section*{Methods}
We consider a dielectric medium with laser pump radiation $(\omega_1, \ \textbf{k}_1)$ propagating inside it. The electrostriction effect causes a medium to undergo periodic deformation. Such a deformation is described by phonon wave vector $\textbf{q}_1$ collinear to $\textbf{k}_1$ and phonon frequency $\Omega_1$. This coherent lattice vibration modulates the material's dielectric constant, thereby causing the scattering of the pump light. The scattering satisfies the phase matching condition \cite{Boyd}:

\begin{equation}
\textbf{k}_2  = \textbf{k}_1 - \textbf{q}_1 ,\quad \omega_2 = \omega_1 - \Omega_1
\label{eq:bragg_first}
\end{equation}
Here $\textbf{k}_2$ and $\omega_2$ are the wave vector and frequency of scattered Stokes light. In a medium with linear dispersion for both photon and phonon, Eq.~(\ref{eq:bragg_first}) yields $\Omega_1=2nv/c \times\omega_1$ \cite{gundavarapu2019sub}, where $v$ is the speed of sound in the medium, $n$ is the refractive index.

The generated Stokes wave $(\omega_2, \mathbf{k}_2)$ can, in turn, act as a pump for a scattering process, inducing a new acoustic wave $(\Omega_2, \mathbf{q}_2)$. This process can continue further, giving rise to more Stokes waves with the number limited by pump power. Therefore each Stokes $j$-th order originates from Stokes of order $j-1$ scattered on a distinct phonon mode with frequency $\Omega_j$. These frequencies are not independent but are determined by the phase matching conditions:
\begin{equation}
\textbf{k}_{j+1}  = \textbf{k}_j - \textbf{q}_j, \quad \omega_{j+1} = \omega_j - \Omega_j
\label{eq:bragg}
\end{equation}

Usually, each scattering order requires a separate phonon mode, as described by Eq.~(\ref{eq:bragg}). In contrast, here, we consider a system where one forward propagating phonon can facilitate many orders of scattering events. We choose relative phonon damping rate $\Gamma_j/\Omega_j$ such that it is large compared to the relative Brillouin frequency shift $\Omega_j/\omega_j$. On top of this, in order for wavevector ($k$) condition in Eq.~(\ref{eq:bragg}) to be satisfied for a large number of pairs of the optical modes, we consider a system in which the interaction between optical and phononic modes occurs in a single specially engineered segment of the ring resonator. Although limiting the interaction region in this manner increases the threshold, it permits the generation regime considered below. A GeSbS ring resonator similar to that in Ref.~\cite{song2024gesbs} provides a basis for a device with such a single active segment. The length of this segment determines how many transitions between photon modes can be mediated via one pair of phonons. Following the approach of Ref.~\cite{sipe2016hamiltonian}, we estimate the strength of interaction between photon and phonon modes as
\begin{equation}
 |g_j|\simeq g_{ring}\sqrt{\frac{L_{seg}}{L_{ring}}}
 \left|\operatorname{sinc}\left(\frac{\Delta q_j L_{seg}}{2}\right)\right|
 \label{eq:segment_coupling}
\end{equation}
where $\Delta q_j=k_j-k_{j+1}-q_{1,2}$ ($1$ for odd $j$, $2$ for even $j$), $L_{seg}$ is the segment length, and $L_{ring}$ is the ring length, $g_{ring}$ is photon-phonon coupling constant for a hypothetical system where Brillouin-active region covers the entire ring resonator. Here the optical modes are normalized over the ring and the effective phonon is normalized within the segment. 

\begin{figure}[h]
    \centering
    \includegraphics[width=1\linewidth]{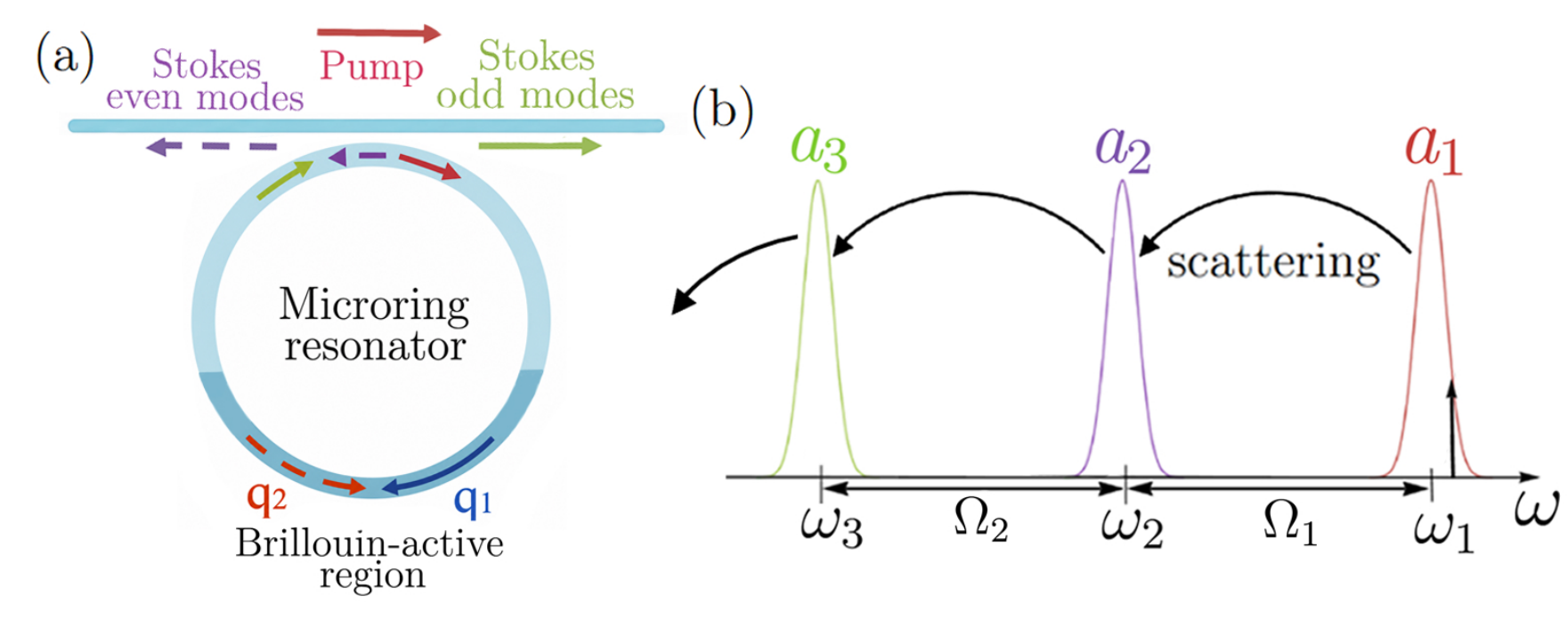}
     \caption{Schematic of the two-phonon-mode Brillouin system in a ring resonator. (a) Microring resonator with propagation directions. Clockwise pump (red, $a_1$) generates clockwise acoustic wave $b_1$ (blue), scattering pump into counter-clockwise first-order Stokes (purple, $a_2$). This Stokes wave generates counter-clockwise acoustic wave $b_2$ (orange), scattering it into clockwise second-order Stokes (green, $a_3$). The cascade continues with odd modes (green) propagating clockwise and even modes (purple) counter-clockwise. The phonon modes are localized in a short suspended segment of the ring. Outside of this segment, acoustic gain is suppressed. (b) Resulting spectrum with approximately equally spaced lines (within the phonon linewidth) at $\omega_j = \omega_1 - \sum_{m=1}^{j-1} \Omega_{s(m)}$, where $s(m)=1$ for odd $m$ and $s(m)=2$ for even $m$.}
    \label{fig:general}
\end{figure}

The Hamiltonian of the described system is as follows:

\begin{equation}
\begin{gathered}
\mathcal H_s = \hbar \sum_{j=1}^{N} \omega_j \hat a_j^\dagger \hat a_j
+ \hbar \Omega_1 \hat b_1^\dagger \hat b_1 + \hbar \Omega_2 \hat b_2^\dagger \hat b_2 \\
- \hbar \sum_{\substack{j=1 \\ j\ \mathrm{odd}}}^{N-1}
(g_j \hat a_j^\dagger \hat a_{j+1} \hat b_1 + g_j^* \hat a_{j+1}^\dagger \hat a_j \hat b_1^\dagger) \\
- \hbar \sum_{\substack{j=2 \\ j\ \mathrm{even}}}^{N-1}
(g_j \hat a_j^\dagger \hat a_{j+1} \hat b_2 + g_j^* \hat a_{j+1}^\dagger \hat a_j \hat b_2^\dagger) \\
+ \hbar (\mathcal E e^{-i\omega_1 t} \hat a_1^\dagger + \mathcal E^* e^{i\omega_1 t} \hat a_1).
\end{gathered}
\label{eq:hamiltonian}
\end{equation}
Here $\hat{a}_j$ and $\hat{a}_j^\dagger$ are the annihilation and creation operators of the $j$-th optical mode. $\hat{b}_{1,2}$ and $\hat{b}_{1,2}^\dagger$ are the annihilation and creation operators of the first and second phonon modes, respectively. $g_1$ is the coupling rate between an odd mode and the next even mode, $g_2$ is the coupling rate between an even mode and the next odd mode. $\mathcal E$ is the amplitude of the external drive, which excites the first optical mode $a_1$. In this notation, even $j$ corresponds to backward propagation, and odd $j$ corresponds to forward propagation. $N$ is the number of the photon modes. The general scheme of a scattering process is presented in Fig.~\ref{fig:general}. To demonstrate the two-phonon cascade, we choose $N=9$.

The interaction terms in Eq.~(\ref{eq:hamiltonian}) are written in the energy-conserving (rotating-wave) form: the term $g_1\hat{a}_j^\dagger\hat{a}_{j+1}\hat{b}_1$ annihilates a lower-frequency Stokes photon and a phonon while creating a higher-frequency pump photon, satisfying $\omega_j = \omega_{j+1}+\Omega_1$.

Using the Heisenberg-Langevin approach \cite{scully1997quantum,dutra2005cavity,hard2024}, we obtain the following equations for the slowly varying amplitudes at resonance, including damping and delta-correlated Markovian noise:

\begin{equation}
\label{eq:general}
\begin{aligned}
\frac{\partial a_1}{\partial t}
&= i g_1 b_1 a_2 - \frac{\gamma_1}{2} a_1 - i\mathcal E, \\[5pt]
\frac{\partial a_{2j}}{\partial t}
&= i g_{2j+1} b_2 a_{2j+1} + i g_{2j}^* b_1^* a_{2j-1} - \frac{\gamma_{2j}}{2} a_{2j}, \\[5pt]
\frac{\partial a_{2j+1}}{\partial t}
&= i g_{2j}^* b_2^* a_{2j} + i g_{2j+1} b_1 a_{2j+2} - \frac{\gamma_{2j+1}}{2} a_{2j+1}, \\[5pt]
\frac{\partial a_N}{\partial t}
&= i g_{N-1}^* b_2^* a_{N-1} - \frac{\gamma_N}{2} a_N, \\[5pt]
\frac{\partial b_1}{\partial t}
&= i \sum_{\substack{j=1 \\ j\ \mathrm{odd}}}^{N-1} g_j^*a_{j+1}^* a_j - \frac{\Gamma_1}{2} b_1 + \chi_1(t), \\[5pt]
\frac{\partial b_2}{\partial t}
&= i\sum_{\substack{j=2 \\ j\ \mathrm{even}}}^{N-1} g_{j}^*a_{j+1}^* a_j - \frac{\Gamma_2}{2} b_2 + \chi_2(t).
\end{aligned}
\end{equation}

Here $N$ is odd. For even $N$, the equation for the last optical mode becomes $\partial a_N/\partial t = i g_1^* b_1^* a_{N-1} - (\gamma_N/2)a_N$. The equations for $a_{2j}$ and $a_{2j+1}$ apply to the intermediate optical modes. We use a semiclassical description, in which $a_j$ and $b_s$ are complex stochastic amplitudes and products of operators are replaced by products of these amplitudes. The coefficients $\gamma_j$ and $\Gamma_s$ are the photon and phonon decay rates, respectively. The phonon noise sources have zero mean and satisfy $\langle \chi_j^*(t + \tau)\chi_k(t)\rangle = \Gamma_j N_j^{ph}\delta_{jk}\delta(\tau)$, $\langle \chi_j(t + \tau)\chi_k(t)\rangle = 0$, $\langle \chi_k(t)\rangle = 0$. Here $N_j^{ph}$ is the thermal phonon occupation, with $N_j^{ph} \approx 8\times10^2$ at room temperature, and $\delta(\tau)$ is the Dirac delta function.

Equation~(\ref{eq:general}) is written for rotating frame at the optical frequencies $\omega_j$ and the phonon frequencies $\Omega_s$, where $s=1,2$. The laboratory-frame amplitudes are $a_j^{\mathrm{lab}}(t) = a_j(t)e^{-i\omega_j t}$ and $b_s^{\mathrm{lab}}(t) = b_s(t)e^{-i\Omega_s t}$. At resonance, the free oscillation terms cancel in these frames. The drive is resonant with the first optical mode and therefore enters the equation for $a_1$ as the constant term $-i\mathcal E$.
Following \cite{hard2024}, we include noise only in the phonon mode equations due to the negligible contribution from photon noise.

The model parameters we use are: $N=9$ (further increasing the number of modes $N$ has little effect on amplitudes and on the flatness of frequency combs), $L_{ring} \approx 2\,\mathrm{cm}$, $\lambda_1 = 1550\,\mathrm{nm}$, the refractive index of waveguide material $n_g \approx 2.4$, $\Omega_1/(2\pi) = \Omega_2/(2\pi) \approx 8\,\mathrm{GHz}$, $\gamma_j/(2\pi) \approx 200\,\mathrm{MHz}$ for all optical modes. We use $\Gamma_1/(2\pi) = \Gamma_2/(2\pi) \approx 50\,\mathrm{MHz}$ and a coupling rate for the full ring of $g_{ring} \approx 5\times10^4\,\mathrm{s^{-1}}$. Similar optical and Brillouin parameters, allowing coupling rates of this order, can be achieved in GeSbS resonators and waveguides \cite{song2024gesbs,song2021gesbs}. For a single interaction segment of length $L_{seg} \approx 1\,\mathrm{mm}$, we take $g_{ring} = 4.47\times10^{-4}\,\mathrm{s^{-1}}$. $L_{seg}= 1\,\mathrm{mm}$ ensures than all $g_j$ are no less than $0.8g_1$, as described by~\ref{eq:segment_coupling}. For noise calculation, we assume that at $T = 300\,\mathrm{K}$, the phonon mode occupation number is determined by thermal equilibrium, so that $N_j^{ph} \approx 10^3$.

\section*{Results}
\begin{figure}[h]
    \centering
    \includegraphics[width=1\linewidth]{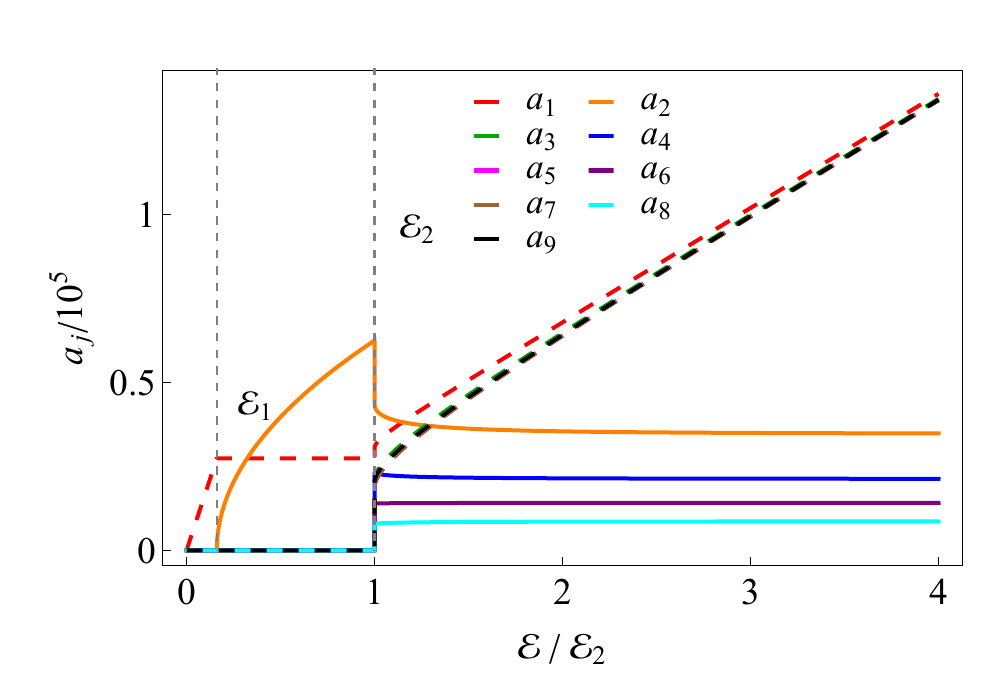}
    \caption{Stationary amplitudes of nine optical modes ($a_j$, $j=1,..,9$) as a function of the normalized pump amplitude $\mathcal E/\mathcal E_2$ up to $4\mathcal E_2$. The two thresholds correspond to the first Stokes mode and to the collective multimode branch. Flatness for odd-numbered modes reaches $9\times10^{-4}$ dB at maximum pump rate.}
    \label{fig:n5plot}
\end{figure}

First, we consider the system without noise terms, $\chi_j=0$. To demonstrate the possibility of the cascade we choose $N = 9$ photonic modes optical modes. The results of simulation are shown in Fig.~\ref{fig:n5plot}. Two notable thresholds in behavior are present. The first threshold corresponds to first Stokes mode excitation, $\mathcal{E}_1$, and the second threshold corresponds to cascade of Stokes modes, $\mathcal{E}_2$. Notably, only two thresholds appear in the output radiation, in contrast with many-phonon model \cite{behunin2018fundamental}, where each consequent Stokes mode features progressively higher pump threshold $\mathcal{E}_2$. The lower turning point of the multimode stationary branch corresponds to the estimated on-chip pump power $P_{in}(\mathcal E_2)\simeq4.20\,\mathrm{W}$ for the proposed short interaction segment. Fig.~\ref{fig:n5plot} demonstrates highly uniform amplitudes for odd-numbered modes, with discrepancy with the highest amplitude being only 1.002 times higher than the lowest, which translates into $\approx9\times10^{-4}$ dB amplitude flatness or $\approx1.7\times10^{-3}$ dB power flatness.

 To calculate spectra, we consider Eq.~(\ref{eq:general}) with noise terms $\chi_j(t)$. We linearize the equations near the stationary solution and project the phonon noise onto the neutral phase modes to estimate the Stokes linewidths, following Ref.~\cite{behunin2018fundamental}. The broadened Stokes contributions are placed at their optical carrier frequencies and summed separately for odd- and even-numbered modes. For this illustration, we use the phonon occupation corresponding to an effective temperature of $300\,\mathrm{K}$.

\begin{figure}[h]
    \centering
    \includegraphics[width=1\linewidth]{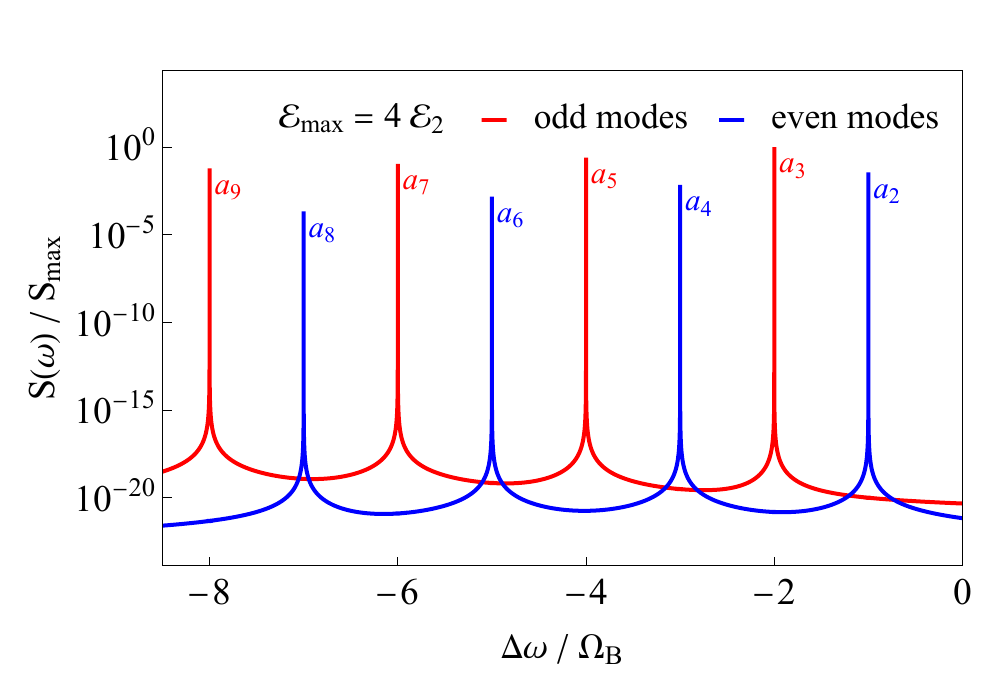}
    \caption{Calculated Stokes spectrum of the cascaded Brillouin scattering at $\mathcal E=4\mathcal E_2$. The normalized optical spectral power density is plotted on a logarithmic scale as a function of the frequency detuning $\Delta\omega$ from the central pump frequency $\omega_1$. The spectrum shows eight phase-diffusion-broadened Stokes lines generated via the two effective phonon modes. Odd-numbered optical modes are shown in red; even-numbered modes are shown in blue. The coherent pump is omitted. Notably, the amplitudes depicted in this figure do not represent total power within respective modes, as peak widths differ. }
    \label{fig:stokes}
\end{figure}     

Notably, Fig.~\ref{fig:n5plot} demonstrates that above the second threshold, $\mathcal{E}_2$, odd numbered optical modes that propagate along the direction of pump mode reach similar amplitudes that increase with pump power. The spectral peaks of such modes have different heights and widths, but show approximately equal integrated powers of the co-propagating Stokes lines, see Fig.~\ref{fig:stokes}.
The even-numbered modes behave differently, there amplitudes tend to constant amplitudes. Thus, coupled with a waveguide, the suggested system enables generation of two combs propagating in opposite directions. The approximately equal integrated powers of the co-propagating Stokes lines are desirable for optical signal processing, in particular, for analogue-to-digital converters \cite{fang2025320}.

\section*{Discussion and Conclusion}
In conclusion, we show that comb generation based on backward Brillouin scattering is possible in a system with only two phonon modes. We find that since only two phonon modes are responsible for drive of Stokes modes, there are two pump thresholds in output curve. The first threshold corresponds to excitation of forward propagating phonons and produces first Stokes mode. The second threshold corresponds to excitation of counter propagating phonons. Above the second threshold, where both types of phonons are present, many orders of Stokes modes are generated simultaneously. In this case all odd Stokes modes are driven by the phonons propagating along pump wave, and all even modes are driven by the phonons propagating in the opposite direction. At high pump rates, the system produces approximately equal integrated powers of the co-propagating Stokes lines, a useful property for optical signal processing.

The model of two phonon modes is applicable to systems where phonon lifetime is short, such that the width of a single phonon mode relative to phonon frequency ($\Gamma_j /\Omega_{j} $) is significantly larger than Brillouin shift (which is usually equal to free spectral range of the photon cavity) relative to photon frequency ($\Omega_{j}/\omega_j $). This alone, however, is insufficient. Due to phonons and photons coexisting in the same geometry, mode overlap integrated over the resonator will only leave transitions where one phonon facilitates one optical transition (phase mismatch will be equal to one period, hence integral will be zero). Therefore, in lifetime requirement, to fulfill the wavevector conditions, it is necessary to limit the region where phonons and photons interact. In this case the same phonon mode can enable Brillouin scattering between multiple pairs of optical modes. Ultimately, this means that all phonons propagating in one direction can be approximately described as one effective phonon with amplitude being the sum of amplitudes of original phonons. Thus, two effective phonons (propagating clockwise and counterclockwise) are sufficient.  The parameters used in our model correspond to $\Omega_j/\omega_j \approx 4\times10^{-5}$ and $\Gamma_j/\Omega_j \approx 6\times10^{-3}$. Appropriate optical and Brillouin parameters can be found in GeSbS resonators \cite{song2024gesbs,song2021gesbs} or $\text{SiN}$ platforms where $\Gamma_j /\Omega_{j} \approx 10^{-2}$, and $\Omega_{j}/\omega_j \approx 2 vn/c $ is as low as $10^{-6}$ \cite{lu2016stimulated, gundavarapu2019sub}. For the proposed single active segment, $P_{in}=\hbar\omega_1|\mathcal E|^2/\gamma_c$ with $\gamma_c=\gamma_1(1-Q_L/Q_i)$ (where $Q_L, Q_i$ are loaded and intrinsic quality factors, respectively), so the estimated threshold input power at which cascade appears is $P_{in}(\mathcal E_2) \simeq 4.2\,\mathrm{W}$. Pulsed pumping could supply this peak power with lower average power \cite{lafforgue2020broadband}. Thus a ring with one Brillouin-active segment can in principle realize the proposed system.


We only consider nine optical modes in our simulation. Further cascading will be impeded by wavevector shift between consequent optical modes deviating from the phonon frequency by more than inverse phonon spatial coherence length, causing non-resonant interaction with the same effective phonon. In practice this will result in excitation of new phonon modes, that will enable further transitions. However, this is a distinct mechanism described in much detail in other works \cite{behunin2018fundamental,wang2025cascading} that shows dissimilar behavior to the two-phonon Brillouin cascade studied in this work. Cascading on further phonon modes can be prevented by use of phonon bandgap structures \cite{safavi2014two,li2024study}. If further cascade is suppressed in such a manner, only two desirable phonon modes will be excited, and the system behaves as depicted in Figure~\ref{fig:n5plot}.

A frequency comb with approximately equal integrated powers of the co-propagating Stokes lines can be used for information processing, in particular, for spectrally sliced analogue-to-digital conversion \cite{fang2025320, zazzi2020fundamental}. This technique uses individual peak amplitudes in a frequency comb to store information. For this purpose, approximately equal integrated powers of the co-propagating Stokes lines are desirable to avoid preliminary filtering. Here, we have demonstrated that this can alternatively be achieved via Brillouin scattering on short-lived phonons. Notably, this feature is not present in Brillouin cascade on narrow phonons, where each consequent optical mode amplitude is lower than the previous one \cite{behunin2018fundamental}. Another advantage of the suggested method is that Brillouin scattering does not require anomalous dispersion in the medium.

Existing methods of comb generation with normal dispersion materials usually involve dark solitons \cite{he2016dynamics, campbell2024frequency, wang2023flat} and electro-optical modulation \cite{mohammadi2025high, kumar2023generation}. These alone are not inherently uniform and typically feature roll-off or triangular spectra for dark solitons  \cite{he2016dynamics, campbell2024frequency, wang2023flat}. Therefore, additional filtering and/or careful engineering of parameters is required to suppress amplitude variation, allowing 0.7-1 dB flatness for several comb lines in some electro-optical modulation setups \cite{zeng2026integrated}. Better flatness or larger bandwidth have only been achieved in complicated setups with additional stabilization and careful second and third order dispersion engineering \cite{li2025ultra}. The method we suggest that is based on two broad phonon scattering provides inherently uniform combs that consists of a single pumped Brillouin cascade laser that does not require any further manipulations. We believe that the suggested method of comb generation via two phonon Brillouin cascade can provide an alternative route to achieving uniform combs without filtering, potentially with sub 0.002 dB flatness, distinct from dark soliton combs and combs via electro-optic modulation.

\section*{Acknowledgment}
E.R.V., I.V.D., A.A.Z. and E.S.A. thank the foundation for the advancement of theoretical physics and mathematics “Basis”.

\bibliography{sample}

\end{document}